\documentclass[twocolumn,amssymb, nobibnotes, showpacs, superscriptaddress, aps, prd]{revtex4}
\usepackage{amsmath}
\usepackage{epsfig}
\usepackage{graphicx}
\usepackage{subfigure}
\usepackage{dcolumn}
\usepackage{bm}

\usepackage[bookmarks=true,
   colorlinks=true,
   linkcolor=blue,
   urlcolor=blue,
   citecolor=blue,
   bookmarks=true,
   hyperindex=true
]{hyperref}
\usepackage{graphicx} 
\usepackage{epstopdf}
\usepackage{siunitx}
\usepackage{bookmark}
\usepackage{amsmath}

\begin{document}

\title{Self-Alignment of a Large-Area Dual-Atom-Interferometer Gyroscope Using Parameter Decoupled Phase Seeking Calibrations}

\author{Zhan-Wei Yao}
\affiliation{State Key Laboratory of Magnetic Resonance and Atomic and Molecular Physics, Innovation Academy for Precision Measurement Science and Technology, Chinese Academy of Sciences, Wuhan 430071, China}
\affiliation{Center for Cold Atom Physics, Chinese Academy of Sciences, Wuhan 430071, China}
\author{Hong-Hui Chen}
\affiliation{State Key Laboratory of Magnetic Resonance and Atomic and Molecular Physics, Innovation Academy for Precision Measurement Science and Technology, Chinese Academy of Sciences, Wuhan 430071, China}
\affiliation{School of Physics, University of Chinese Academy of Sciences, Beijing 100049, China}
\author{Si-Bin Lu}
\affiliation{State Key Laboratory of Magnetic Resonance and Atomic and Molecular Physics, Innovation Academy for Precision Measurement Science and Technology, Chinese Academy of Sciences, Wuhan 430071, China}
\affiliation{Center for Cold Atom Physics, Chinese Academy of Sciences, Wuhan 430071, China}
\author{Run-Bing Li}
\email[]{rbli@wipm.ac.cn}
\affiliation{State Key Laboratory of Magnetic Resonance and Atomic and Molecular Physics, Innovation Academy for Precision Measurement Science and Technology, Chinese Academy of Sciences, Wuhan 430071, China}
\affiliation{Center for Cold Atom Physics, Chinese Academy of Sciences, Wuhan 430071, China}
\author{Ze-Xi Lu}
\affiliation{State Key Laboratory of Magnetic Resonance and Atomic and Molecular Physics, Innovation Academy for Precision Measurement Science and Technology, Chinese Academy of Sciences, Wuhan 430071, China}
\affiliation{School of Physics, University of Chinese Academy of Sciences, Beijing 100049, China}
\author{Xiao-Li Chen}
\affiliation{State Key Laboratory of Magnetic Resonance and Atomic and Molecular Physics, Innovation Academy for Precision Measurement Science and Technology, Chinese Academy of Sciences, Wuhan 430071, China}
\affiliation{School of Physics, University of Chinese Academy of Sciences, Beijing 100049, China}
\author{Geng-Hua Yu}
\affiliation{State Key Laboratory of Magnetic Resonance and Atomic and Molecular Physics, Innovation Academy for Precision Measurement Science and Technology, Chinese Academy of Sciences, Wuhan 430071, China}
\affiliation{Center for Cold Atom Physics, Chinese Academy of Sciences, Wuhan 430071, China}
\author{Min Jiang}
\affiliation{State Key Laboratory of Magnetic Resonance and Atomic and Molecular Physics, Innovation Academy for Precision Measurement Science and Technology, Chinese Academy of Sciences, Wuhan 430071, China}
\affiliation{Center for Cold Atom Physics, Chinese Academy of Sciences, Wuhan 430071, China}
\author{Chuan Sun}
\affiliation{State Key Laboratory of Magnetic Resonance and Atomic and Molecular Physics, Innovation Academy for Precision Measurement Science and Technology, Chinese Academy of Sciences, Wuhan 430071, China}
\affiliation{School of Physics, University of Chinese Academy of Sciences, Beijing 100049, China}
\author{Wei-Tou Ni}
\affiliation{State Key Laboratory of Magnetic Resonance and Atomic and Molecular Physics, Innovation Academy for Precision Measurement Science and Technology, Chinese Academy of Sciences, Wuhan 430071, China}
\author{Jin Wang}
\affiliation{State Key Laboratory of Magnetic Resonance and Atomic and Molecular Physics, Innovation Academy for Precision Measurement Science and Technology, Chinese Academy of Sciences, Wuhan 430071, China}
\affiliation{Center for Cold Atom Physics, Chinese Academy of Sciences, Wuhan 430071, China}
\author{Ming-Sheng Zhan}
\email[]{mszhan@wipm.ac.cn}
\affiliation{State Key Laboratory of Magnetic Resonance and Atomic and Molecular Physics, Innovation Academy for Precision Measurement Science and Technology, Chinese Academy of Sciences, Wuhan 430071, China}
\affiliation{Center for Cold Atom Physics, Chinese Academy of Sciences, Wuhan 430071, China}

\date{\today}

\begin{abstract}
We realize a Mach-Zehnder-type dual-atom-interferometer gyroscope with an interrogation arm of 40 cm length and the interference area up to 1.2 cm$^2$. The precise angular alignment of the large-scale separated Raman lasers is demonstrated by seeking the phase intersection of Ramsey-Bord$\acute{e}$ interferometers after the gravity effect is compensated and by decoupling the velocity dependent crosstalk phase shifts, and applied to build the Mach-Zehnder atom interferometer. Then a compact inertial rotation sensor is realized based on dual large-area Mach-Zehnder atom interferometers by precisely aligning the large-scale separated Raman lasers, in which the coherence is well preserved and the common noise is differentially suppressed. The sensor presents a sensitivity of $1.5\times10^{-7}$ rad/s/Hz$^{1/2}$, and a stability of $9.5\times10^{-10}$ rad/s at 23000 s. The absolute rotation measurement is carried out by adjusting the atomic velocity which corresponds to modulating the scale factor.
\end{abstract}

\pacs{06.30.Gv, 37.25.+k, 03.75.Dg, 32.80.Qk}

\maketitle

High-precision gyroscope plays an important role in fields of inertial navigation\cite{Lefevre2013a}, fundamental physics\cite{Ciufolini2004a} and geophysics\cite{Brosche1998a}. The studies on atom interferometers, in past two decades, make it possible to develop an atom-interferometer gyroscope with high rotation sensitivity and stability\cite{Durfee2006a,Gauguet2009a,Berg2015a,Yao2018a,Savoie2018a,Dutta2016a,Stockton2011a,Xu2020a}.  Two types of atom-interferometer gyroscopes with Mach-Zehnder (M-Z) and four-pulse configurations were developed, and their features have been discussed\cite{Barrett2014a}. In principle, the M-Z interferometer is suitable for the wider environments including the weightless condition in space, and is much beneficial for building a spatial high-sensitivity gyroscope due to the fact that the common-mode noise can be differentially suppressed in the dual atom interferometers\cite{Jentsch2004a,Canuel2006a}. The sensitivity of an atom-interferometer gyroscope, based on Sagnac effect, scales with the atomic interference-loop area. Building the large-area atom-interference loop is an important avenue for improving the sensitivity of the atomic gyroscope. To enlarge the interference-loop area, large-momentum-transfer techniques, including Raman adiabatic passage\cite{Kotru2015a}, Bragg diffractions\cite{Chiow2011a} and Bloch oscillation\cite{Muller2009a}, were developed. However, these techniques are still in the infant hood, and their feasibilities are not well verified in atom-interferometer gyroscopes.

\vskip 3pt
\noindent

As an effective and practical method for improving the sensitivity, the interference-loop area can be enlarged by synchronously increasing the interrogation arm and the atomic longitudinal velocity in the M-Z-type atom-interferometer gyroscope. However, as the interrogation arm lengthened, to build a closed interferometric loop, the separated Raman beams should be more precisely aligned for combining two interferometric paths within the coherent length of atomic wave packets. Using self-alignment of separated Raman lasers with the contrast of Ramsey-Bord$\acute{e}$ (R-B) interferometers, the interference-loop area was enlarged to 0.2 cm$^2$ in the M-Z atom interferometer\cite{Tackmann2012a,Yao2018a}, but which is still smaller than one in the four-pulse atom interferometer\cite{Savoie2018a,Xu2020a}. Here, to build a larger-area M-Z-type atom interferometer, the coherent manipulation of atomic wave packets will be limited by the angular misalignment of larger-scale separated Raman lasers. The angular alignment of separated Raman lasers faces a huge challenge and is still an obstacle for building the larger-area M-Z atom interferometer. Compared with the contrast, the phase of the R-B interferometer is more sensitive to the angular alignment of separated Raman lasers. It was applied to analyze the systematic error in the four-pulse atom fountain interferometer, where two Raman lasers are aligned at the 0.2 $\mu$rad level\cite{Altorio2020a}. While in the M-Z atom interferometer, to align separated Raman lasers, the separated time ($T_{1}$) should be modulated, which makes the precision for seeking the phase intersected point seriously limited by the cross-coupled parameters including the gravity effect and the velocity dependent crosstalk phase shift\cite{Tackmann2012a}. Thus it is urgently needed to develop an effective method for the precise angular alignment of large-scale separated Raman lasers, because of which is a key step for building the larger-area M-Z-type atom-interferometer gyroscope.

\vskip 3pt
\noindent

In this letter, we demonstrate an effective method for precisely aligning the large-scale separated Raman lasers using parameter decoupled phase seeking calibrations, and present a high-precision inertial rotation sensor with dual large-area M-Z-type atom interferometers. First, the parameter decoupled phase seeking calibration is performed by seeking the phase intersected point of symmetric R-B interferometers after the gravity effect is compensated and by decoupling the velocity dependent crosstalk phase shifts. Three large-scale separated Raman lasers with an interrogation arm of 40 cm are precisely aligned at the 0.4 $\mu$rad level. Then, dual M-Z atom interferometers are realized with the interference-loop area of 1.2 cm$^2$, which is the largest loop area in the M-Z-type atom interferometer to our knowledge. Finally, a high-precision inertial rotation sensor is realized based on dual large-area atom interferometers, where the dead time is reduced and the common-mode noise is differentially canceled. The performance is evaluated by Allan deviations and it reaches a stability of $9.5\times10^{-10}$ rad/s after integrating for 23000 s. The absolute rotation measurement is demonstrated by modulating the atomic velocities.

\begin{figure}[htp]
	\centering
		\includegraphics[width=0.48\textwidth]{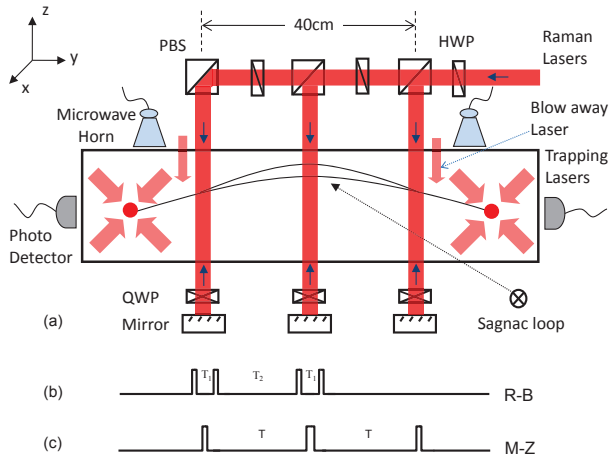}
	\caption{(color online)  Schematic diagram of the compact gyroscope with dual large-area cold-atom interferometers (a), timing sequences of Ramsey-Bord$\acute{e}$ (b) and Mach-Zehnder interferometers (c). The length of vacuum chamber is less than 80 cm. The atoms trapped in two MOTs are launched and then propagated in the vacuum chamber with opposite directions along the symmetrical parabolic trajectories ($y$-$z$ plane). Three pairs of separated Raman lasers, which are aligned by adjusting angles of mirrors ($x$/$y$-axes), are counter-propagated along the gravity direction ($z$-axis), and two symmetric large-area atom-interferometer loops are built for measuring the horizontal rotation component. $T_{1}$ is the separated time between the first/last two pulses, and $T_{2}$ is the delay time between the second and third pulses in R-B interferometers. $T$ is the interrogation time between two adjacent Raman pulses in M-Z interferometers. PBS: polarizing beam splitter, HWP: half wave plate, and QWP: quarter wave plate. }%
	\label{fig-1}%
\end{figure}

\vskip 3pt
\noindent

The schematic apparatus is shown in Fig.\textit{\ref{fig-1}} (a). We build a compact inertial sensor with dual cold atom interferometers. The length of the sensor is 80 cm, in which the atomic propagating path is 56 cm and the interrogation arm is extended to 40 cm. Cold atoms are counter-propagating in the x-y plane, with a slight angle along the y-axis direction. The dead time caused by the non-interference time during the atom propagating process, which limits the sensitivity of the atom-interferometer gyroscope\cite{Biedermann2013a,Yao2016a}, is reduced because the same vacuum window is shared between the trapping light and the Raman lasers. The inner height of the vacuum chamber is more than 5 cm, which makes it possible to extend the interrogation time up to $T=$50 ms. The laser system is similar to that in our previous work\cite{Zhang2018a}, and all optical elements are integrated into a compact module with dimensions of 50 cm$\times$50 cm$\times$16 cm. The laser frequencies are tuned by acoustic-optic modulators (AOMs) to match the energy levels of  $^{87}$Rb atoms. The Raman lasers are delivered to the interferometric area by a polarized-maintained optical fiber. They are collimated by an achromatic doublet lens with a focal length of 150 mm, and split into three pairs of Raman beams by polarized beam splitters and half wave plates. Three tip/tilt platforms driven by the piezoelectric ceramic (PZT), which can provide an angular resolution to less than sub-microradians, are used to fix three mirrors.

\vskip 3pt
\noindent

The $^{87}$Rb atoms are loaded into two magneto-optical traps (MOTs) from the background vapor in 0.5 s, and then are launched by the moving optical molasses technique at a slight angle with respect to the horizontal direction. When considering the gravity effect, the atomic trajectory has a position displacement along the gravity direction during their propagations along the horizontal direction. The atomic longitudinal velocities can be adjusted from 3.6 to 10 m/s by controlling the detuning of trapping lasers. When the atoms are accelerated from the MOTs, the polarized gradient cooling is applied in a few milliseconds. After the atoms are prepared to the magnetic insensitive state of $\vert$F=1,m$_F$=0$\rangle$ by a microwave field and the residual atoms on the other states are blown away by the resonance lasers. The initial atoms are manipulated by three pairs of separated Raman lasers along the gravity direction. The geomagnetic field is compensated by three pairs of Helmholtz coils and the ac Stark shift is canceled by adjusting the intensity ratio of Raman lasers\cite{Li2008a,Li2009a}. The population of the state $\vert$F=2,m$_F$=0$\rangle$ is detected by the laser induced fluorescence signal, and the atomic interference fringes are observed by scanning the differential phase of the third pair of Raman lasers.

\vskip 3pt
\noindent

To construct a closed interferometric loop, the separated Raman beams should be precisely aligned for combining two interferometric paths within the coherent length of atomic wave packets. The alignment of Raman beams was realized by the contrast of symmetric R-B interferometers\cite{Yao2018a,Tackmann2012a}, while it was less stringent on the alignment precision. Here, the key avenue to realize a M-Z interferometer is how to transit from two symmetric R-B interferometers to one M-Z interferometer. However, as the interrogation arms elongated, the larger size disparity between the Raman laser and the interrogation arms make it more difficult to improve alignment precision when transiting from the two symmetric R-B interferometers with two pairs of Raman beams to one M-Z interferometer with three pairs of Raman beams\cite{Tackmann2012a,Yao2018a}. Compared to the contrast measurement, the phase measurement provides a higher sensitivity\cite{Xu2017a,Altorio2020a}, which is suitable to the precise angular alignment in the four-pulse interferometer. For the symmetric R-B interferometer, the phase imprinted by Raman lasers is written as
\begin{equation}\label{eq1}
\phi_{RB}=\phi_{1}(\vec{r}_{1})-\phi_{2}(\vec{r}_{2})-\phi_{3}(\vec{r}_{3})+\phi_{4}(\vec{r}_{4})
\end{equation}
where, $\phi_{i}=\vec{k}_{\rm eff} \cdot \vec{r}_{i}$ ($i=1,2,3,4$) is the phase imprinted by Raman lasers on the atoms, $k_{\rm eff}$ is the effective wave vector of Raman lasers, and $\vec{r}_{i}$ ($x,y,z$) is the spatial position of the atoms interacted with the Raman lasers. The tilt of Raman lasers will change the phase shift when the Raman lasers interacted with the atoms, which can be measured by symmetric R-B interferometers. As the Raman beams are tilted in a tiny range, considering the first order approximation, the phase is given by
\begin{equation}\label{eq3}
\phi_{RB}=k_{\rm eff} [v_{x} \theta_{x}+v_{y} \theta_{y}-(g-\frac{\alpha}{k_{\rm eff}})(T_{1}+T_{2})]T_{1}
\end{equation}
where, $g$ is the gravity acceleration, and $\alpha$ is the chirped rate. $v_{j}$ ($j=x,y$) is the velocity component, and $\theta_{j}$ is the relative angle between two pairs of Raman lasers on the $x$- and $y$-axis directions, respectively. $T_{1}$ is the separated time between the first/last two pulses, and $T_{2}$ is the delay time between the second and third pulses, as shown in Fig.\textit{\ref{fig-1}} (b). The phase is sensitive to the separated time, and the phase intersection of R-B interferometers with the different separated times could give a criterion on whether two separated Raman lasers are parallel. However, in the experiment, it is very difficult to get the phase intersected point due to the velocity dependent crosstalk phase shifts when the separated time is modulated\cite{Tackmann2012a}.

\vskip 3pt
\noindent

\begin{figure}[htp]
	\centering
		\includegraphics[width=0.48\textwidth]{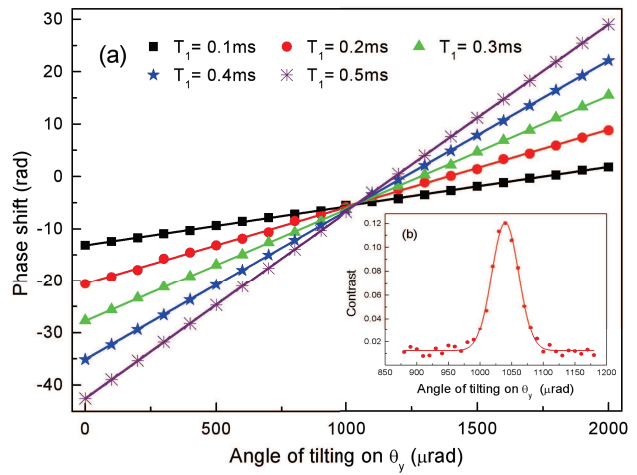}
	\caption{(color online) (a) Phase dependence of Ramsey-Bord$\acute{e}$ atom interferometers on the angular tilting of Raman lasers with different separated times of $T_{1}$=0.1 ms (black squares), 0.2 ms (red dots), 0.3 ms (green triangles), 0.4 ms (blue stars) and 0.5 ms (purple diamonds). The experiment results are linearly fitted, and the intercross point is an absolute zero-phase point when two pairs of Raman lasers are parallel. (b) Contrast dependence of Mach-Zehnder interferometers on the angular tilting of Raman lasers with the interrogation time of $T$=50 ms. The result is fitted by the Gaussian function.}%
	\label{fig-2}%
\end{figure}

Here we demonstrate an effective method for seeking the phase intersected point of symmetric R-B interferometers by decoupling the gravity-induced and velocity-dependent crosstalk phase shift. To improve the alignment accuracy, we introduce a chirped rate $\alpha$ into the Raman lasers to compensate the gravity caused phase shift. The deviation of $\alpha$ induces a systematic error in the angular alignment. If we want to seek the phase intersection, the uncertainty of chirped rate should be confined to a range of a few kHz/s, which corresponds to the 10$^{-4}$ level of the gravity measurement.
To accurately compensate the chirped rate, we construct a M-Z interferometer with the second pair of Raman lasers to measure the gravity compensated chirp rate\cite{Zhou2011a}. Limited by the beam size, the maximum interrogation time is about 2.6 ms, which can give a chirped rate of 25.104 $\pm$ 0.001 MHz/s. The uncertainty of the chirped rate induces a deviation of few microradians in the alignment, which is sufficient to compensate the gravity-caused phase shift.
The phase dependence on the tilting angles is measured in the symmetric R-B interferometer after the gravity is compensated by applying a chirped rate. We first measure the contrast of symmetric R-B interferometers to get an angular resolution of one hundred of microradians. Then an optical slit is added on the detector to select the atomic velocity, which reduces the demand of angular alignment by several times. With the same delay time ($T_{2}=50$ ms) and different separated times ($T_{1}$) from 0.1 to 0.5 ms, the phase dependence of symmetric R-B interferometer on $\theta_{y}$ is shown in Fig.\textit{\ref{fig-2}} (a). The systematic error, caused by the wavefront distortion, is ignored for the short separated time. The phase of the R-B interferometer depends linearly on $\theta_{y}$, and due to the much less transverse velocity component, the phase is not sensitive to $\theta_{x}$. The velocity dependent crosstalk phase shift between the $x$- and $y$-axis directions is much reduced due to their velocity components decoupled. Thus the phase seeking calibration is carried out by compensating the gravity effect and by decoupling the velocity dependent crosstalk phase shift. There are several phase intersected point for different separated times, which provides a criterion for precisely seeking the angular alignment of separated Raman lasers. The relative angular of two Raman lasers on the $y$-axis direction is measured and given by the standard deviation, and it is $1039.69 \pm 10.60$ $\mu$rad from Fig.\textit{\ref{fig-2}} (a). For the larger separated time, the phase is more sensitive to the angular alignment. This is a key step to build the large-area M-Z atom interferometer.

\vskip 3pt
\noindent

Three pairs of Raman beams are precisely aligned with the method mentioned above, and applied to build dual large-area M-Z interferometers as shown in Fig.\textit{\ref{fig-1}} (c). For the inertial measurement, we use the velocity sensitive pulses to manipulate atoms. The different pulse durations select the atoms of different velocity distributions. To reduce the contrast loss caused by the different durations, the intensity ratio of three pairs of Raman beams are set to $1:2:1$, which provides the ideal durations for selecting the atoms with the same velocity group. Then, the optical slit is removed, and the interference fringe is observed in the M-Z interferometer. To maximize the contrast of fringes, three pairs of separated Raman beams with a consecutive distance of 20 cm should be more precisely aligned. The alignment of three Raman lasers is optimized by adjusting the atomic trajectories and the chirped rate, which makes the velocity component perpendicular to the propagating direction of Raman lasers tended to zero. In this case, the velocity-dependent crosstalk phase shift is further reduced, and the angular alignment the separated Raman lasers is improved by decoupling the velocity dependent crosstalk phase shifts on $\theta_{x}$ and $\theta_{y}$. The contrast of M-Z fringes is measured by tilting the angle on $\theta_{y}$, and is fitted by the Gaussian function, as show in Fig.\textit{\ref{fig-2}} (b). The contrast of the fringe is up to 12$\%$ using three pairs of separated Raman beams with a consecutive time of 50 ms. The angular alignment dependent on $\theta_{y}$ is measured at the 0.4 $\mu$rad level by the contrast of M-Z interferometers. In the dual atom interferometers, the trajectory difference makes the atoms experience the different transition probability, which reduces the fringe contrast\cite{Yao2018a}. To maximize the common noise rejection ratio, a trade-off between two interferometers is realized by optimizing the atomic trajectories and the Raman pulses. Finally, we realized a dual M-Z-type cold-atom-interferometer gyroscope with contrasts of 7.6$\%$ and 8.4$\%$, which are limited by the overlapping of velocity-sensitive Raman transitions.

\begin{figure}[htp]
	\centering
		\includegraphics[width=0.48\textwidth]{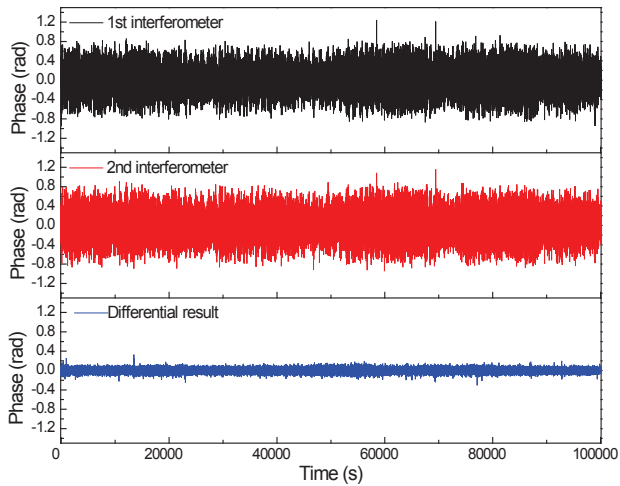}
	\caption{(color online) Phases of the dual atom interferometers are measured by using the square modulation method. Although the phases are fluctuated for the 1st interferometer (black curve) and the 2nd interferometer (red curve), their phase noise is canceled in the differential mode (blue curve).}%
	\label{fig-3}%
\end{figure}

\vskip 3pt
\noindent

\begin{figure}[htp]
	\centering
		\includegraphics[width=0.48\textwidth]{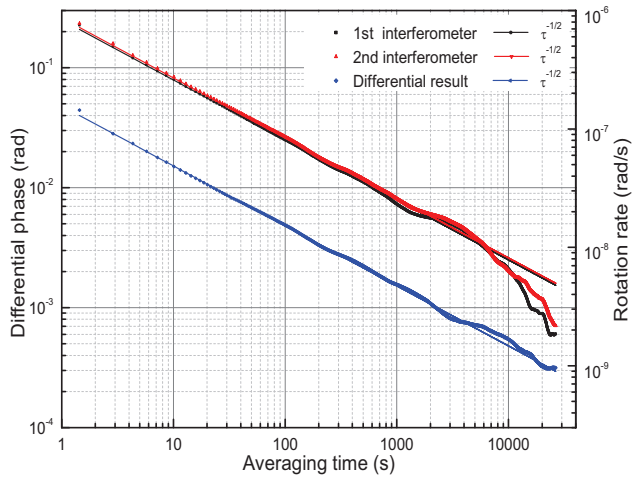}
	\caption{(color online) Performance is evaluated by the Allan deviations. The differential result of dual atom interferometers shows a common-noise rejection in the gyroscope.}%
	\label{fig-4}%
\end{figure}

The performance of the atomic gyroscope is evaluated with the same method as used in the previous work\cite{Yao2018a}. The phases of dual atom interferometers are measured by using the square modulation method as shown in Fig.\textit{\ref{fig-3}}. The signals are set to two adjacent mid-fringe points by adjusting the phase difference of the Raman lasers. To improve rejection effect of common-mode noise, we make the two fringes in phase by modulating the interference-loop area. In this case, the atomic longitudinal velocity is 4.1 m/s, and the interference area is 1.2 cm$^2$, which is the largest area reported in M-Z-type interferometers. The dominant noise term originates from the vibration noise, which is estimated by measuring the environment vibration and calculating its contribution using the weighting function\cite{Cheinet2008a}. Fortunately, the vibration noise can be rejected in dual atom interferometers. Besides of the short-term noise, the thermal expansion of the platform changes the optical path length among three pairs of Raman beams, which produces a phase shift. As the interrogation arm elongated, this effect is more serious. The long-term drift is also a common-mode phase and eliminated in dual atom interferometers. To eliminate the long-term drift, we add a feedback loop to compensate the temperature-caused phase drifts, which acts as a constant phase for the dual interferometers, and can be subtracted as a common-mode phase during the rotation measurement. The above results are evaluated by using the Allan deviations, as shown in Fig.\textit{\ref{fig-4}}. The black squares are phases for the 1st atom interferometer, and the red triangles are those for the second one. The blue diamonds are the differential results for the dual-atom interferometers. The black, red and blue lines show the corresponding ideal white noise curves, as a slope of $\tau$$^{-1/2}$ in the double logarithmic coordinate system. Due to the loop feedback, for the single interferometer, the signal deviates from $\tau$$^{-1/2}$ when the integrating time is longer than 6000 s. The phase uncertainty after common noise rejection using two interferometers demonstrates a 5-time noise compression compared with the single atom interferometer, approaching to 40 mrad per shot at the integrating time of 23000 s. Considered with the cycle time of 1.44 s, the short-term rotation sensitivity is $1.5\times10^{-7}$ rad/s/Hz$^{1/2}$. Allan standard deviations decrease with the integration time as $\tau$$^{-1/2}$ after several seconds, reaching a long-term stability of $9.5\times10^{-10}$ rad/s at 23000 s. To our knowledge, this performance is the state of art of the reported M-Z-type atom-interferometer gyroscope.

\vskip 3pt
\noindent

The absolute rotation measurement is important for an inertial sensor. The phase of an atom interferometer is usually a periodic quantity, which introduces an extra phase shift of $2m\pi$  ($m$ is an integer number). This becomes more obvious for the large-area atom interferometer. In our experiment, even for the Earth' rotation rate, the phase shift of atom interferometer is on several tens of radians. To realize an absolute rotation measurement, we measure the Earth' rotation rate with different scale factors. The scale factor is modulated by adjusting the atomic longitudinal velocity, which is different from varying the interrogation time in the atom gravimeter\cite{Wu2019a,Bonnin2018a,Lautier2014a,Avinadav2019a}. The absolute rotation rate is measured by extracting the differential phases for different scale factors. The Earth' rotation is measured and its rate is $(6.067 \pm 0.006) \times 10^{-5}$ rad/s. The horizontal component of Earth' rotation rate is $6.280 \times 10^{-5}$ rad/s when considering the latitude of $30.55^{\circ}$ in Wuhan. This deviation is verified by using a commercial compass. It is caused by the misalignment between the interference area and the latitude direction.

\vskip 3pt
\noindent

In conclusion, the atom-interferometer gyroscope with the high rotation sensitivity and long-term stability has many potential applications in scientific and technical fields. The dual M-Z-type atom-interferometer gyroscope is much suitable to three-axis rotation measurements. We realized a compact M-Z-type large-area atom-interferometer gyroscope by precisely aligning the large-scale separated Raman lasers, in which the dead time is reduced as much as possible by elongating the interrogation arms. More importantly, the parameter decoupled phase seeking calibration is proposed and performed for aligning the separated Raman lasers. Three Raman lasers are aligned at the 0.4 $\mu$rad level.
For the larger separated time, the phase measurement is more sensitive to the tilting angular alignment, which can provide an effective method for aligning the larger-scale separated Raman lasers by more precisely compensating the gravity effect and further decoupling crosstalk parameters dependent phase shift. It will be useful for realizing the larger-area interferometer, such as a large-scale atom-interferometer gyroscope\cite{Zhan2019a}, where the angular alignment is more urgent. After precisely aligning the Raman lasers, a high-precision gyroscope is realized using dual M-Z-type atom interferometers with each atomic interference-loop area of 1.2 cm$^2$, and the absolute rotation rate is measured by modulating the atomic velocities. With the dual interferometers for rejecting the common-mode noise, the sensor reaches a long-term stability of $9.5\times10^{-10}$ rad/s after integrating for 23000 s. In the future, the rotation sensitivity could be further improved by reducing the dead time using the continuous operation\cite{Meunier2014a} and by suppressing the amplitude noise using the normalized detection\cite{Biedermann2009a}.
\vskip 3pt
\noindent

We acknowledge the financial support from the National Key Search and Development Program of China under Grant No. 2016YFA0302002, the National Natural Science Foundation of China under Grant Nos. 11674362, 91536221, and 91736311, the Strategic Priority Research Program of Chinese Academy of Sciences under Grant No. XDB21010100, the Outstanding Youth Foundation of Hubei Province of China under Grant No. 2018CFA082, the National Defense Science and Technology Innovation Project of China, and the Youth Innovation Promotion Association of Chinese Academy of Sciences.

\bibliographystyle{plain}
\bibliography{ref}

\bibliography{basename of .bib file}

\begin{thebibliography}{99}

\bibitem {Lefevre2013a} H. C. Lef$\grave{e}$vre, The fiber-optic gyroscope: Challenges to become the ultimate rotation-sensing technology, Opt. Fib. Techn.  \textbf {19}, 828 (2013).
\bibitem {Ciufolini2004a} I. Ciufolini, and E. C. Pavlis, A confirmation of the general relativistic prediction of the Lense-Thirring effect, Nature \textbf {431}, 958 (2004).
\bibitem {Brosche1998a} P. Brosche, and H. Schuh, Tides and Earth rotation, Sur. Geophys. \textbf {19}, 417 (1998).
\bibitem {Durfee2006a} D. S. Durfee, Y. K. Shaham, and M. A. Kasevich, Long-term stability of an area-reversible atom-interferometer Sagnac gyroscope, Phys. Rev. Lett. \textbf {97}, 240801 (2006).
\bibitem {Berg2015a} P. Berg, S. Abend, G. Tackmann, C. Schubert, E. Giese, W. P. Schleich, F. A. Narducci, W. Ertmer, and E. M. Rasel, Composite-light-pulse technique for high-precision atom interferometry, Phys. Rev. Lett. \textbf {114}, 063002 (2015).
\bibitem {Dutta2016a} I. Dutta, D. Savoie, B. Fang, B. Venon, C.L. Garrido Alzar, R. Geiger, and A. Landragin, Continuous Cold-Atom Inertial Sensor with 1 nrad/sec Rotation Stability, Phys. Rev. Lett. \textbf {116}, 183003 (2016).
\bibitem {Stockton2011a} J. K. Stockton, K. Takase, and M. A. Kasevich, Absolute Geodetic Rotation Measurement Using Atom Interferometry, Phys. Rev. Lett. \textbf {107}, 133001 (2011).
\bibitem {Gauguet2009a} A. Gauguet, B. Canuel, T. L$\acute{e}$v$\grave{e}$que, W. Chaibi, and A. Landragin, Characterization and limits of a cold-atom Sagnac interferometer, Phys. Rev. A \textbf {80}, 063604 (2009).
\bibitem {Yao2018a} Z. W. Yao, S. B. Lu, R. B. Li, J. Luo, J. Wang, and M. S. Zhan, Calibration of atomic trajectories in a large-area dual-atom-interferometer gyroscope, Phys. Rev. A \textbf {97}, 013620 (2018).
\bibitem {Savoie2018a} D. Savoie, M. Altorio, and B. Fang, L. A. Sidorenkov, R. Geiger, A. Landragin, Interleaved atom interferometry for high-sensitivity inertial measurements, Sci. Advan. \textbf {4}, eaau7948 (2018).
\bibitem {Xu2020a} W. J. Xu, L. Cheng, J. Liu, C. Zhang, K. Zhang, Y. Cheng, Z. Gao, L. S. Cao, X. C. Duan, M. K. Zhou, and Z. K. Hu, Effects of wave-front tilt and air density fluctuations in a sensitive atom interferometry gyroscope, Opt. Express \textbf {28}, 12189 (2020).
\bibitem {Barrett2014a} B. Barrett, R. Geiger, I. Dutta, M. Meunier, B. Canuel, A. Gauguet, P. Bouyer, A. Landragin, The Sagnac effect: 20 years of development in matter-wave interferometry, C. R. Physique \textbf {15}, 875 (2014).
\bibitem {Jentsch2004a} C. Jentsch, T. M$\ddot{u}$ller, E. M. Rasel, HYPER: A Satellite Mission in Fundamental Physics Based on High Precision Atom Interferometry, Gen. Rel. Grav. \textbf {36}, 2197 (2004).
\bibitem {Canuel2006a} B. Canuel, F. Leduc, D. Holleville, A. Gauguet, J. Fils, A. Virdis, A. Clairon, N. Dimarcq, Ch. J. Bord$\acute{e}$, A. Landragin and P. Bouyer, Six-axis inertial sensor using cold-atom interferometry, Phys. Rev. Lett. \textbf {97}, 010402 (2006).
\bibitem {Kotru2015a} K. Kotru, D. L. Butts, J. M. Kinast, and R. E. Stoner, Large-Area Atom Interferometry with Frequency-Swept Raman Adiabatic Passage, Phys. Rev. Lett. \textbf {115}, 103001 (2015).
\bibitem {Chiow2011a} S. W. Chiow, T. Kovachy, H. C. Chien, and Mark A. Kasevich, 102$\hbar$k Large Area Atom Interferometers, Phys. Rev. Lett. \textbf {107}, 130403 (2011).
\bibitem {Muller2009a} H. M$\ddot{u}$ller, S. W. Chiow, S. Herrmann, and and S. Chu, Atom Interferometers with Scalable Enclosed Area, Phys. Rev. Lett. \textbf {102}, 240403 (2009).

\bibitem {Tackmann2012a} G. Tackmann, P. Berg, C. Schubert, S. Abend, M. Gilowski, W. Ertmer and E. M. Rasel, Self-alignment of a compact large-area atomic Sagnac interferometer, New J. Phys. \textbf {14}, 015002 (2012).
\bibitem {Altorio2020a} M. Altorio, L. A. Sidorenkov, R. Gautier, Accurate trajectory alignment in cold-atom interferometers with separated laser beams, Phys. Rev. A  \textbf {101}, 033606 (2020).

\bibitem {Biedermann2013a} G. W. Biedermann, K. Takase, X. Wu, L. Deslauriers, S. Roy, and M. A. Kasevich, Zero-Dead-Time Operation of Interleaved Atomic Clocks, Phys. Rev. Lett. \textbf {111}, 170802 (2013).

\bibitem {Yao2016a} Z. W. Yao, S. B. Lu, R. B. Li, L. Cao, J. Wang, and M. S. Zhan, Continuous dynamic rotation measurements using a compact cold atom gyroscope, Chin. Phys. Lett. \textbf {33}, 083701 (2016).

\bibitem {Zhang2018a} X. W. Zhang, J. Q. Zhong, B. Tang, X. Chen, L. Zhu, P. W. Huang, J. Wang, and M. S. Zhan, Compact portable laser system for mobile cold atom gravimeters, Appl. Opt. \textbf {57}, 6545 (2018).
\bibitem {Li2009a} R. B. Li, P. Wang, H. Yan, J. Wang, and M. S. Zhan, Magnetic field dependence of coherent population transfer by the stimulated Raman transition, Phys. Rev. A  \textbf {77}, 033425 (2008).
\bibitem {Li2008a} R. B. Li, L. Zhou, J. Wang, and M. S. Zhan, Measurement of the quadratic Zeeman shift of $^{85}$Rb hyperfine sublevels using stimulated Raman transitions, Opt. Commun. \textbf {282}, 1340 (2009).
\bibitem {Xu2017a} W. J. Xu, M. K. Zhou, M. M. Zhao, K. Zhang, and Z. K. Hu, Quantum tiltmeter with atom interferometry, Phys. Rev. A  \textbf {96}, 063606 (2017).

\bibitem {Zhou2011a} L. Zhou, Z. Y. Xiong, W. Yang, B. Tang, W. C. Peng, Y. B. Wang, P. Xu, J. Wang, and M. S. Zhan, Measurement of local gravity via a cold atom interferometer, Chin. Phys. Lett. \textbf {28}, 013701 (2011).
\bibitem {Cheinet2008a} P. Cheinet, B. Canuel, F. P. D. Santos,  A. Gauguet, F. Yver-Leduc, and A. Landragin, Measurement of the sensitivity function in a time-domain atomic interferometer, IEEE Trans. Instrum. Meas. \textbf {57}, 1141 (2008).
\bibitem {Wu2019a} X. Wu, Z. Pagel, B. S. Malek, T. H. Nguyen, F. Zi, D. S. Scheirer, and H. M$\ddot{u}$ller, Gravity surveys using a mobile atom interferometer, Science Adv. \textbf {5}, eaax0800 (2019).
\bibitem {Bonnin2018a} A. Bonnin, C. Diboune, N. Zahzam, Y. Bidel, M. Cadoret, and A. Bresson, New concepts of inertial measurements with multi-species atom interferometry, Appl. Phys. B \textbf {124}, 180 (2018).
\bibitem {Lautier2014a} J. Lautier, L. Volodimer, T. Hardin, S. Merlet, M. Lours, F. P. D. Santos, and A. Landragin, Hybridizing matter-wave and classical accelerometers, Appl. Phys. Lett. \textbf {105}, 144102 (2014).
\bibitem {Avinadav2019a} C. Avinadav, D. Yankelev, O. Firstenberg, and N. Davidson, Composite-fringe atom interferometry for high dynamic-range sensing, arXiv:1912.12304v1.
\bibitem {Zhan2019a} M. S. Zhan, J. Wang, W.T. Ni, D. F. Gao, G. Wang, L. X. He, R. B. Li, L, Zhou, X. Chen, J. Q. Zhong, B. Tang, Z. W. Yao, L. Zhu, Z. Y. Xiong, S. B. Lu, G. H. Yu, Q. F. Cheng, M. Liu, Y. R. Liang, P. Xu, X. D. He, M. Ke, Z. Tan, and J. Luo, ZAIGA: Zhaoshan Long-baseline Atom Interferometer Gravitation Antenna, Inter. J. Mod. Phys. D. \textbf {29}, 1940005 (2020).
\bibitem {Meunier2014a} M. Meunier, I. Dutta, R. Geiger, C. Guerlin, C. L. Garrido Alzar, and A. Landragin, Stability enhancement by joint phase measurements in a single cold atomic fountain, Phys. Rev. A \textbf {90}, 063633 (2014).

\bibitem {Biedermann2009a} G. W. Biedermann, X. Wu, L. Deslauriers, K. Takase, and M. A. Kasevich, Low-noise simultaneous fluorescence detection of two atomic states,  Opt. Lett. \textbf {34}, 347 (2009).

\end{thebibliography}

\end{document}